\documentclass[12pt]{article}
\usepackage{fullpage}

\def\to{\rightarrow}
\begin{document}

\begin{center}
\Large{\bf QUANTUM MECHANICS AND REALITY$^\star$} \\
\bigskip\bigskip
\large{Virendra Singh} \\
\bigskip
Tata Institute of Fundamental Research \\
Homi Bhabha Road, Colaba, Bombay 400 005 
\end{center}
 
\bigskip\bigskip\bigskip
\begin{center}
\underbar{Abstract} \\
\end{center}
\medskip

We begin by discussing ``What exists?'', i.e. ontology, in Classical
Physics which provided a description of physical phenomena at the
macroscopic level.  The microworld however necessitates a introduction
of Quantum ideas for its understanding.  It is almost certain that the
world is quantum mechanical at both microscopic as well as at
macroscopic level.  The problem of ontology of a Quantum world is a
difficult one.  It also depends on which interpretation is used.  We
first discuss some interpretations in which Quantum Mechanics does
not provide a complete framework but has to be supplemented by extra
ingredients e.g. (i) Copenhagen group of interpretations associated
with the names of Niels Bohr, Heisenberg, von-Neumann, and (ii)
de-Broglie-Bohm interpretations.  We then look at some 
interpretations in which Quantum mechanics is supposed to provide the
entire framework such as (i) Everett-deWitt many world, (ii) quantum
histories interpretations.  We conclude with some remarks on the
rigidity of the formalism of quantum mechanics, which is sharp
contrast to it's ontological fluidity.

\vfill

\noindent $^\star$Invited talk at Workshop on ``Foundations of Sciences'',
National Institute of Advanced Study, Bangalore (23-25 February 2004).

\begin{verbatim} 
[e-mail: vsingh@theory.tifr.res.in]
\end{verbatim}

\newpage

\hrule width0pt

\vfill

\newpage

\noindent {\large\bf 0. Introduction}
\medskip

Classical Physics is a wonderful and satisfying theoretical edifice.
It is a closed theory which has hardly any epistemological problem
and it comes with a simple ontology of particles and waves (or fields).
It is also empirically successful in a large domain of physics.
Around the end of ninteenth century there began to appear some clouds
on the horizon suggesting its' empirical inadequateness.  These had
mainly to do with problem of blackbody radiation, atomic structure and
energy-exchanges between radiation and atomic systems.  These led to
the introduction of quantum ideas in 1900 by Planck.  After an
exploratory phase of old quantum theory during the first quarter of
twentieth century.  The final mathematical formulation of Quantum
mechanics was achieved by Heisenberg (1925), Schr\"odinger (1926) and
Dirac (1925).

The change in our picture of physical reality, ushered by quantum
revolution, is extremely drastic and revolutionary.  To put this
change in perspective we shall first briefly recapitulate the picture
of physical reality (ie ontology) in Classical Physics.  We shall then
exhibit three phenomenon to illustrate the weird nature of quantum reality.
After a brief enunciation of the ``rules of the game'' i.e. the
framework of quantum physics, we discuss the nature of reality in
Quantum physics.  The discussion depends on whether, while
interpreting the formalism of quantum mechanics, it is to be
supplemented by extra ingredients or not.  The examples of the first
kind is provided by (i) various type of Copenhagen interpretations and
(ii) de-Broglie Bohm interpretation.  The examples of interpretations,
when quantum mechanical formalism is all there is, are (i) Many world
interpretation and (ii) Quantum History approach.  We close with a few
concluding remarks on the contrast between rigidity of the formalism
and the fluidity of the ontology of quantum mechanics.
\bigskip

\noindent {\large\bf 1. Clasical Physics and Reality}
\medskip

The edifice of Classical Physics was created by Issac Newton in his
book ``Philosophiae Naturalis Principia Mathematica'' (1687).  This 
incorporated earlier insights obtained by N. Copernicus, J. Kepler,
G. Galileo and other giants on whose shoulders he stood.  Classical
Physics was enriched by the work of Faraday and Maxwell on the
electromagnetism in mid-nineteenth century, and by the special theory
of relativity (1905) and general theory of relativity (1915) of Albert
Einstein. 
\medskip

\noindent {\large 1.1. Newtonian Dynamics:}
\medskip

Newtonian world consist of discrete mass-points, each with a given
mass, which move along their trajectories in a three dimensional space
with time.  They obey Newton's three laws of motion and influence each
others' motion through a force exerted by them on each other.  As
Newtonian laws of motion are differential equation in time for the
mass-point positions, the dynamics is deterministic.  Further since
the differential equations are of second order in time, a specification of
the initial conditions, i.e. the position and velocity of all the
mass-point at the initial time, fixes the entire trajectory of motion
for all times.  The ontology of the Newtonian mechanics is thus that
of a large number of particles, with given masses, moving along their
trajectories. 

Newton also discovered the universal law of gravitation: ``any two
mass points attract each other with a force proportional to the
product of their masses and inversely proportional to the square of
the distance between them.  When we discuss the motion of
``macroscopic'' bodies then gravitation is the only relevant force.
It applies equally to the fall of apples on the earth, tides, shape of
the earth, the motion of moon around the earth, the motion of planets
around the Sun.  Newton was thus able to show that Celestial and
terrestrial phenomenon are described by the same physics.  In short
astronomy is a branch of physics.

The various forces between the particles, in general are supposed to
decrease with increasing distance between them as exemplified by
Newton's law of gravitation.  Otherwise independant particle concept
becomes ambiguous.  Further note that Newtonian forces are an example of
``action at a distance'' i.e. a particle affects another particle at a
distance.  It is not a contact interaction.

In Newtonian ontology there are only particles and their
trajectories.  The continuum approximation to Newtonian dynamics of a
large number of particles results in phenomenological theories such as
hydrodynamics, theory of elasticity.  These phenomenological theories
do describe observation such as sound waves, water waves etc.  So the
waves exist as an `epiphenomena'.

We also note down, for future reference, that Newtonian conception of
space is that of an three dimensional Euclidean space.  The time is
taken to one dimensional straight line.  The space and time provide a
fixed arena in which particle move but which are themselves not
affected by any motion of the particles.  We thus have an absolute
space and absolute time. 
\medskip

\noindent {\large 1.2. Faraday-Maxwell Electromagnetic Field Theory:}
\medskip

The Danish physicist H.C. Oersted was the first one to demonstrate a
relationship between electricity and magnetism.  He found in 1820 an
electric current flowing in a wire deflects a magnetic needle placed
parallel to it.  Ampere, as a result of his work on forces during
1822-27 exerted by two current carrying wires on each other, showed
that an electric current is equivalent to a magnetic shell in its
effects.  Ampere proposed his hypothesis that all magnetism is really
due to small electric current loops inside the magnetic materials.
The converse effect, ie whether a magnet can behave like an electric
current, was looked for by Faraday.  During those investigations, he
discovered his laws of electromagnetic induction in 1831.  In order to
understand these laws of electromagnetic induction, he was led to
introduce the concept of a physical electromagnetic field.  He found
the clue in the old practice of strewing iron fillings around a bar
magnet to exhibit the effects of a magnetic force.  The iron filing
arrange themselves in closed curves.  To Faraday these curves
suggested that idea that the whole of space is filled with these
invisible lines, indicating the direction of magnetic forces.  Further
he assumed that their density is indicative of the strength of the
magnetic force.  A similar picture applies for the electric force.  He
thus pictured electric and a magnetic field existing at all the space
points, for all time, surrounding electric charges and magnets.  In
view of their interrelationship it is better to call this new physical
entity an electromagnetic field.

Maxwell, begining with his paper ``On Faraday's Lines of Force''
(1856), investigated the problem of physical origin of electromagnetic
field and the precise mathematical formulation of laws of
electromagnetic field.  He also introduced the concept of
``displacement current''.  It was initially done on the analogy of
displacements induced in a dielectric medium but later abstracted to
be there simply by virtue of presence of an electric field.  It was
found that electromagnetic field is an ontologically new physical
entity.  It is not reducible to any rotating cylinders, mechanical
gears or any such devices.

The final mathematical formulation of electromagnetic field theory now
known as Maxwell's Equation, was achieved by Maxwell in 1864.

In Newtonian universe force between two electric charges, given by
Coulomb's law, was taken to be as a ``action at a distance''.  With the
field picture this can be interpreted as a ``local interaction''.  An
electric charge is surrounded by an electric field and it is this
electric field which affects another charge.  This can be seen to lead
to observed Coulomb's law of force between.  As Faraday said ``Matter
cannot act where it is not''.  One can reinterpret even Newton's Law
of gravitational force as a ``local interaction'', rather than
``action at a distance'', by introducing a gravitational field around
a mass point.

A remarkable consequence of Maxwell's equation was the prediction of
electromagnetic waves with a fixed velocity, $c$, which agreed with
the known velocity of light.  It was thus possible to regard light as
an electromagnetic radiation.  It is through Faraday-Maxwell's
electromagnetic field that ``waves'' appear as ontologically
fundamental entities in classical physics for the first time.  In
talking of Maxwell's theory Boltzmann often waxed lyrical and quoted
G\"oethe's Faust (Part 2),

\begin{enumerate}
\item[{}] ``Was it a god, who wrote these \\ which with mysterious,
hidden means \\ the forces of nature about me unveil \\ and fill my heart
with peaceful joy''.
\end{enumerate}
\bigskip

\noindent {\large 1.3. Relativity theory of Einstein:}
\medskip

One of the presuppositions of physics of the nineteenth century was
that we need a physical medium for any wave propagation.  A luminiferous 
aether was duly postulated as the medium for the propagation of
electromagnetic waves.  The laws of Newtonian dynamics are invariant
under Galilean transformations corresponding to a uniform rectilinear
motion between two physical systems while the laws of electromagnetism
are not so.  It should therefore be possible to determine the velocity
between say earth and the aether.  Repeated attempts were unsuccessful
in determining this velocity.  The problem was resolved finally by
Einstein through his special theory of relativity (1905).  In special
theory of relativity, Newtonian equations of mechanics are modified so
as to invariant under the same symmetry.  group of transformations as
Maxwell's equations of electromagnetism viz Lorentz group.  The space
and time are now no longer separate entities and absolute but merge
into other.  As Minkowski put it ``The space and time, as separate
existence, are now doomed to oblivion ....''.$^2$  And further it was
realised that there need be no aether.  Electromagnetic waves are
perfectly happy to propagate in vacuum.

In General theory of Relativity Einstein further modified the
background Minkowskian space-time to a Riemannian curved space time.
The presence of Gravitational force was seen to be equivalent to
curvature effects of space-time.  Gravitation field was thus seen to
be same as metric-tensor of curved space time.  Matter thus reacts
upon the space-time causing it to develop curvature.  Space-time is
not just an area on which particles and fields play, it is also an
actor in the play.
\bigskip

\noindent {\large 1.4. Ontology of Classical Physics:}
\medskip

\noindent {\it 1.4.1. Laws and Real External World}
\medskip

In his delightful little book ``Nature and the Greeks'' Schr\"odinger
emphasises two basic hypotheses which Greek thinkers made about
Nature.  The first of these refers to ``existence of a real External
World''.  This amounts to taking the observer, or rather his
consciousness, out of the description of the observed world.  The
observer is external to the world.  The physical bodies of the
observer are, of course, part of this real external world.  The second
basic hypothesis is that this external world is comprehensible through
existence of laws of nature.  That this is a special hypothesis about
the real external world is clear from it's comparison with the private
world of dreams for which such laws are presumably not there.  These
are also the foundations on which classical physics is based.

Another important feature of classical physics is it's unitary
nature.  It describes both the system under observation and the
measuring apparatus as well as their mutual interaction.  Because of
interaction every measuring apparatus does disturb the system under
observation but one can reduce this disturbance to any desired level
by using gentler and gentler probes.  Classical physics describes the
world as it is without any split between the observed system and
observing apparatus.  Measurement is not an epistomological problem in
classical physics.  It is only a practical problem.
\bigskip

\noindent {\it 1.4.2. Entities}
\medskip

\nobreak
The basic ontological objects of classical physics are point-particles
and fields.  The fields exhibit sometime wave behavior.  These
entities move in a four dimensional space-time continuum; space having
three dimensions and time being a one-dimensional continuum.
\bigskip

\noindent {\it 1.4.3. Determinism}
\medskip

\nobreak
The laws of classical physics are deterministic.  If we specify
initial conditions for a Newtonian universe, consisting of
point-particles interacting with specified forces, then we can predict
it's entire future and retrodict it's entire past.  This, of course,
would require absolute accuracy in the knowledge of initial
conditions.  In practice there may be limitations to our ability to
predict the future and retrodict the past arising out of errors of
observations needed to specify the initial conditions.  It is known
now that some physical systems, i.e. chaotic systems, may have extra
sensitive dependence on initial conditions.  This difficulty can be
tackled by making multiple observations of the system at various
times.  These multiple observations help us reduce the errors in our
initial condition measurements and thus can increase the duration of
time for which useful, i.e. to a specified accuracy, predictions can
be made.

Even in a Newtonian Universe it may not be either practical or
worth-while to predict the behaviour of a classical system when the
total number of degrees of freedom is very large.  It is impractical
because of complexity and amount of computation involved and is not
worthwhile as our interest may be only in a few physical variables,
such as pressure, energy, entropy etc. for a gas in a volume.  We
would not know what to do with the information on the trajectones of
all the particles in a system even if it was available.  In such
situations one of course uses statistical mechanics methods and
probabilities.  The assumption of random behaviour involved in such
considerations is not basic to classical physics.  It is due to
practical exigencies.

The electromagnetic fields and gravitational field obey coupled
partial differential equation which are of first order in time.  They
are thus fully determined once the initial condition for fields,
together with necessary boundary conditions, are specified.

Classical physics is thus intrinsically deterministic and causal.
\bigskip

\noindent {\it 1.4.4. Locality} 
\medskip

\nobreak
We have already mentioned how the observer has been abstracted out of
the total system.  The success of classical physics is based on
another successful hypothesis namely that of locality.  It is possible
to deal meaningfully with parts of the external world.  It is not
necessary to deal with all of it at the same time.  To any given
accuracy we can identify independent subsystems of the universe and
deal with them only and forget the rest.  Two such sub-systems which
are far apart do not affect each other appreciably.  We do not keep
worrying while dealing with terrestrial phenomena as to what is
happening in Andromeda galaxy.
\bigskip

\noindent {\large\bf 2. Quantum Phenomena : Three Exhibits}
\medskip

\nobreak
The weirdness of quantum phenomena, in contrast to straightforward
ontology of classical phenomena, will now be exhibited in three
cases.  They will bring home vividly the problem which we face in
working out an ontology of a quantum world.
\bigskip

\noindent {\large 2.1. Double Slit Experiment}
\medskip

\nobreak
In order to bring out the contrast between classical and quantum
physics it is instructive to contrast the behaviour of classical and
quantum objects in a similar situation.  Feynman has emphasized the
gedanken experiment involving the double-slit setup in this
connection. As he says ``we choose to examine a phenomenon which is
impossible, \underbar{absolutely} impossible, to explain in any
classical way, and which has in it the heart of quantum mechanics.  In
reality, it contains the \underbar{only} mystery''.  In this setup
there is a source $S$ in front of a screen $A$ with two parallel slits
$A_1$ and $A_2$ in horizontal direction, and finally another screen
$M$ parallel to it and behind it some distance away and equipped with
proper devices to measure the pattern.
\bigskip

\noindent {\it 2.1.1. Particles.} Consider the case when the source
$S$ shoots \underbar{indivisible} bullets, exemplifying point
particles is classical physics, and we measure the vertical
distribution $P$ of the fraction of the total number of bullets
arriving at $M$.  In this case each bullet will be definitely going
through either slit $A_1$ or through slit $A_2$ and arrive at a
definite point on the screen $M$.  Let the vertical distribution of
bullets observed at $M$ be $P_1,P_2$ and $P_{12}$ respectively
depending on whether only the slit $A_1$, only the slit $A_2$ or both
the slits $A_1$ and $A_2$ are open.  We will find that $P_{12} = P_1 +
P_2$.
\bigskip

\noindent {\it 2.1.2. Waves.} Let us now consider the case when the
source $S$ produces water waves to exemplify the wave behaviour in
classical physics.  We will now measure the vertical distribution $I$
of the intensity of the wave-motion.  We firstly note that now there
is no discreteness as in the case of bullets where the bullet either
arrives fully at a point on screen $M$ or not at all.  The wave
amplitude is continuous.  Further a wave, unlike a bullet, does not go
through either slit $A_1$ or slit $A_2$ if both slits are open.
Denoting $I$ by $I_1,I_2$ and $I_{12}$, for the three situation of (i)
only slit $A_1$, (ii) only slit $A_2$, and (iii) both slits $A_1$ and
$A_2$ open, we will now find that $I_{12} \neq I_1 + I_2$.  The
intensity pattern thus exhibits interference.  The observed situation
can be described in terms of a complex wave amplitude $a$, related to
intensity by $I = |a|^2$ and such that $I = |a_{12}|^2 = |a_1 + a_2|^2
= I_1 + I_2 + 2\sqrt{I_1 I_2} \cos \delta$ where $\delta$ is the relative
phase difference between $a_1$ and $a_2$.  The last term is the
interference term.
\bigskip

\noindent {\it 2.1.3. Electrons.} We now would like to contrast these
cases with that obtained for electrons which will exemplify the
quantum situation.  The source $S$ is now an electron gun.  If we use
a counter at screen $M$ we will find discrete clicks due to electrons
arriving there one by one.  Electrons, like bullets, exhibit
discreteness.  They register either fully or not at all. That is like
particles.  We thus expect for the vertical distribution, with the
same notation as for bullets, $P_{12} = P_1 + P_2$.  This, however,
does not work out.  We find that electrons display interference and
there exists a complex wave-amplitude $\psi$ for electrons, with $P =
|\psi|^2$, and we have $\psi_{12} = \psi_1 + \psi_2$.  Electrons thus
behave as waves.  We noted that electrons are always detected whole.
Classically they must have gone through either slit $A_1$ or slit
$A_2$.  So how can they interfere?  We are up against something
weird.  We would therefore like to follow up an electron to find out
in each case through which slit it went.  We could place a detector
$C$ near the slits for that purpose.  We should now be in a position
to see how the interference arises and what do we find?  We now find
$P_{12} = P_1 + P_2$.  If we know through which slit the electron went
there is no interference pattern any more.  If the detector $C$ has
interacted with the electron sufficiently to know its path it destroys
the wave nature of the electron.  Is it possible to devise any
experimental set up in which the electron simultaneously behaves as a
wave and as a particle?  Many such possible set ups were analysed in
the early period of quantum mechanics during dialogues between Bohr
and Einstein.  It was found that it is never possible.  Quantum
indeterminacy especially Heisenberg's uncertainty principle, always
plays an important role in these discussions.  We thus see that
quantum physics must give up the determinism of the classical physics
at least when there are no hidden degrees of freedom.
\bigskip

\noindent {\it 2.1.4. Delayed Choice Experiment}
\medskip

\nobreak
The assertion that ``an electron must have gone either through one or
through the other slit even when one did not find it out'' becomes
untenable in quantum world.  Quantum physics thus does not describe
the world as it is out there.  The reality is circumscribed by the
experimental set up as well.  This is quite unlike classical physics
where the measurement can be as gentle as one wishes and thus can be
used to describe the world \underbar{as it is out there}.  Wheeler has
emphasised this aspect of the quantum physics even more eloquently
through his ``Delayed-Choice Double-Slit Experiment''.  One could use
a pulsed electron source which is so weak that each pulse has at most
one electron.  The setup using fast electronics could be such that the
decision about whether one wants to observe wave or particle aspect of
electrons could be postponed until after electrons have already passed
through the slits.  Thus depending on our choice of the final setup it
would seem possible to choose at a later time to circumscribe what the
electron did at an earlier time.  As Wheeler puts it ``No phenomenon
is a phenomenon until is an observed phenomenon''.

It is probably worthwhile to point out that with the development of
neutron interferometery it is now possible to carry out experiments
which until recently were possible only as gedanken experiments.
\bigskip

\noindent {\large 2.2. Quantum Nonseparability}
\medskip

\nobreak
\noindent {\it 2.2.1. Einstein-Podolsky-Rosen Correlations}
\medskip

\nobreak
In quantum mechanics if two systems have once interacted together and
later separated, no matter how far, they can not any more be assigned
separate state vectors.  Since physical interaction between two very
distant systems is negligible, this situation is very
counter-intuitive.  Schr\"odinger even emphasised this aspect, ``I
would not call that one but rather the characteristic of quantum
mechanics''. 

In a famous paper ``Can Quantum mechanical description of Reality be
considered complete?'' Einstein, Podolsky and Rosen analysed this
paradoxical aspect of Quantum mechanics.  In order to make more
precise what they meant by the title of their paper we quote two of
their definitions :

\begin{enumerate}
\item[{(i)}] A \underbar{necessary} condition for the completeness of
a theory is that ``every element of the physical reality must have a
counterpart in the physical theory''.
\item[{(ii)}] A \underbar{sufficient} condition to identify an element
of physical reality : ``If, without in any way disturbing a system, we
can predict with certainty (i.e. with probability equal to unity) the
value of a physical quantity, then there exists an element of physical
reality corresponding to this physical quantity''.
\end{enumerate}

The result of their consideration in this paper, as restated by
Einstein in 1949, is the E.P.R. theorem : ``The following two
assertions are not compatible with each other :

\begin{enumerate}
\item[{(1)}] the description by means of the $\psi$-function is
\underbar{complete}, 
\item[{(2)}] the real states of spatially separated objects are
independent of each other''.
\end{enumerate}

Einstein had a strong commitment to the second assertion.  As he put
it ``But on one supposition we should, in my opinion, absolutely hold
fast : the real factual situation of the system $S_2$ in independent
of what is done with the system $S_1$ which is spatially separated
from the former''.  This postulate is therefore referred to as
``Einstein locality postulate''.  If quantum mechanics is a complete
description then this eminently plausible postulate must be
incorrect. 

We now describe the EPR gedanken experiment in the version proposed by
Bohm as it avoids using continuum wavefunctions.  We consider a two
particle system, with each particle having spin ${1\over2}$, in a
singlet (i.e. total spin 0) state e.g. postronium.  If this system
decays, and if the decay interaction is such that angular momentum is
conserved, then the spin wavefunction would continue to be that
corresponding to a singlet state. When the particles have separated
sufficiently far we measure say $z$-component of spin $S^{(1)}_z$ of
the first particle.  IF the result of this measurement is
$\pm{1\over2}$ then we can assert that $z$ component $S^{(2)}_z$ of
the second particle is $\mp{1\over2}$.  Since this can be done without
disturbing the second particle the $z$-component of the spin of the
second particle is to be regarded as an element of reality and
existing in the second particle alone.  If so the second particle must
have had the same value of $S_z$ before the measurement.

At this juncture we note that we were equally free to have measured
some other component, say $x$-component, of spin $S^{(1)}_x$ of the
first particle.  The singlet wavefunctions can be equally well
represented using a basis in which $S^{(1)}_x, S^{(2)}_x$ are diagonal
instead of $S^{(1)}_z, S^{(2)}_z$.  By a similar reasoning we can
assert that $S^{(2)}_x$ must also be an element of reality.  The
second particle must have had the same value of $S^{(2)}_x$ before the
measurement of $S^{(1)}_x$.  This is however impossible since the
second particle can not simultaneously have a definite value of both
$S^{(2)}_x$ and $S^{(2)}_z$ as these donot commute.  The EPR theorem
follows. 
\bigskip

\noindent {\it 2.2.2. Bell's Inequalities}
\medskip

\nobreak
The Einstein locality postulate is intrinsially very appealing.  Bell
therefore investigated whether we could preserve this postulate if
hidden variables are there for the system.

Consider again two spin ${1\over2}$ particles in a singlet state and
moving in opposite directions towards two measuring Stern-Gerlach
magnets.  The result of measurement $A(\hat a,\lambda)$ of the $\hat
a$-component of spin of the first particle $\vec \sigma_1 \cdot \hat
a$ would be either $+1$ or $-1$ and can depend on $\hat a$ and the
hidden variables, collectively denoted by $\lambda$, of the system
i.e. $A(\hat a,\lambda) = \pm 1$. Similarly for the result of
measurement $B$ of $\hat b$-component of spin of the second particle
$\vec \sigma_2 \cdot \hat b$ we have $B = B(\hat b,\lambda)$ and $B
(\hat b,\lambda) = \pm 1$.  `Einstein Locality' implies that $A$ does
not depend on $\hat b$ and $B$ does not depend on $\hat a$.  Further
we assume that the measurement epochs are such that no direct light
signal can travel between the two Stern-Gerlach magnets.

We now proceed to measure the spin correlation given by the average
value $P(\hat a,\hat b)$ of the product $AB$.  We can write
\[
P(\hat a,\hat b) = \int d\lambda\rho(\lambda) A(\hat a,\lambda) B(\hat
b,\lambda) 
\]
where $\rho(\lambda)$ is the nonnegative normalized probability
distribution of the hidden variables for the given quantum mechanical
singlet state $\psi$. If we assign, to be even more general, some
hidden variables to the magnets, and average over them also the
expression is somewhat changed to 
\[
P_{h\cdot v} (\hat a,\hat b) = \int d\lambda\rho (\lambda) \bar A
(\hat a, \lambda) \bar B (\hat b, \lambda)
\]
where $|\bar A (\hat a, \lambda)| \leq 1, |\bar B (\hat b,\lambda)|
\leq 1$ the subscript $h \cdot v$ stands for hidden variables. 

An important consequence of the above representation for $P_{h\cdot v}
(\hat a, \hat b)$ is the Bell's inequality, 
\[
|P_{h\cdot v} (\hat a,\hat b) - P_{h\cdot v} (\hat a, \hat b')| +
 |P_{h\cdot v} (\hat a',\hat b') + P_{h\cdot v} (\hat a',\hat b)| \leq
 2. 
\]
We now note that quantum mechanics predicts
\[
P(\hat a,\hat b) = \langle \vec\sigma_1 \cdot \hat a \vec\sigma_2
\cdot \hat b \rangle_{\rm singlet} = - \hat a \cdot \hat b .
\]
This form of $P (\hat a, \hat b)$ cannot satisfy the Bell inequality
for a general choice of directions. It this follows that ``No local
hidden variable theory can reproduce all the results of quantum
mechanics''. 

These inequalities have been derived with weaker assumptions and
further inequalities have also been obtained. Since these inequalities
can be experimentally tested a number of such experiments have been
carried out to decide whether nature respects Einstein
locality. Subtle questions involving efficiencies of detectors are
involved in deciding the outcome but preliminary indications are in
favour of quantum mechanics. 
\bigskip

\noindent {\it 2.2.3. Hardy's version of EPR Correlations}
\medskip

\nobreak
We present it in a paraphrase by Stapp. Consider again two spin
one-half particles, in a correlation quantum state, an moving in
opposite directions towards region A and region B. We can measure two
physical observables for either of two electrons and let us call these
observables by, say ``size'' and ``colour''. The ``size'' can take two
values viz. ``large'' and ``small''. The ``colour'' can be either
``white'' or ``black''. 

Hardy showed that the two particle quantum state can be such that the
following are perfectly correlated i.e. one of them implies the other
with probability equal to one. These are 
\begin{enumerate}
\item[{(i)}] 
If ``size'' were to measured in region A and was found to have the
value ``large'', then if ``colour'' was measured in region B it would
be found to be ``white'' with probality 1. In brief 

[region A : ``size'' = ``large''] $\to$ [region B : ``colour'' =
  ``white''] 
\item[{(ii)}] 
[region B : ``colour'' = ``white''] $\to$ [region A : ``colour'' =
  ``black''] 
\item[{(iii)}] 
[region A : ``colour'' = ``black''] $\to$ [region B : ``size'' =
  ``small'']

Would it be correct to conclude following correlation is also perfect: 

[region A : ``size'' = ``large''] $\to$ [region B : ``size'' =
  ``small'']

as it seems to be implied by the commonsense and classical
physics. Hardy showed that this is not necessarily so in quantum
mechanics. In approximately, upto a maximum of about nine percent of
cases, we will find that [region A : ``size'' = ``large''] would
result in [region B: ``size'' = ``large'']. 
\end{enumerate}

\bigskip

\noindent {\it 2.2.4. Correlations and ``Atomism''}
\medskip

\nobreak
The ``atomism'' is an old doctrine about the world.  It asserts that
the world consists of point-like particles which influence each other
less and less as the interparticle distance increases.  This doctrine,
originating with Democritus in Greece, was also a part of Newtonian
world view.  In India it was held by Vaishesik school of philosophy
founded by Kan$\bar{\rm a}$da.

The Einstein-Podolsky-Rosen correlations in Quantum Mechanics are
nonlocal correlations which do not decrease with distance and violate
Einstein locality.  In ``atomism'' such correlations can arise only
from past common causes.  As discussed Bell showed, and experiments
confirm, that there exist correlations and these can not be understood
in this fashion.  As B. d'Espagnat concludes ``It is now proved that
the world in it's entirety can not be thought of as composed as
modern-time atomism claimed it to be''.
\bigskip

\noindent {\large 2.3. Schr\"odinger's Cat and Wigner's Friend}
\medskip

\nobreak
In order to bring home the counter intuitive nature of the postulate
of collapse of wavefunctions, we describe here the two examples which
go in the literature under the name of Schr\"odinger's cat and
Wigner's friend.

\begin{enumerate}
\item[{(i)}] \underbar{Schr\"odinger's Cat} : Enclose a cat in a steel
chamber provided with the following hellish contraption.  The chamber
has a very tiny amount of a radio active isotope in a Geiger counter.
The amount is so tiny that the probability of an atomic decaying in
one hour is equal to the probability of not decaying.  If the atom
decays the Geiger counter triggers and activates a hammer which breaks
a bottle of cyanide and the cat dies.  If on the other hand the atom
does not decay then the cat is, of course, alive and well.  Thus at
the end of one hour the wavefunction of the cat is a linear
superposition of a pure wavefunction describing a living cat and
another pure wavefunction describing a dead cat i.e. $\psi = {1 \over
\sqrt{2}} \psi$ (cat alive) $+ {1\over\sqrt{2}} \psi$ (cat dead).
Curiosity getting the better of us, at the end of hour, we open the
chamber and look at the cat.  It is found to be either alive or dead.
It is definitely not found in a linear superposition of these two
states.  The wavefunction of the cat has collapsed acausally as a
result of observation.  By looking at it we have either killed the cat
or saved it.  Was the cat really in suspended animation before we
opened the chamber?

If you are not satisfied with the arrangement considered you could
follow Feynmann in dramatising it still further.  The Geiger counter
could be made to trigger a sufficient number of atomic weapons to
destroy the whole earth instead of just killing the cat.

\item[{(ii)}] \underbar{Wigner's Friend} : We imagine a physicist $A$,
who is a firm believer in the collapse of the wavefunction position,
making experimental arrangement to observe a physical system $S$ in a
room.  Let the system $S$ be not in an eigenstate of the observable
which $A$ is going to measure.  As long as $A$ does not make the
required measurement the result ofthe measurement, which is yet to be
made, is uncertain.  Now $A$ completes the measurement resulting in
collapse of wavefunction of $S$ and notes down the result in his
laboratory notebook.

The plot thickens and there is another physicist $B$, who is an
equally orthodox believer, and who is in possession of the wavefunction
of the system $S+A$.  He is interested in the entry in the notebook of
$A$ but not just yet.  After a week from the time of notebook
recording by $A$ the second physicist opens the room and looks at the
notebook.  Being orthodox he belives that his observation has
collapsed the wavefunction of $S+A$ and made the entries in the
notebook of $A$ definite.  Until he opened the door the wavefunction
of the notebook must have been a linear superposition of different
entries.  $B$ therefore, assuming a superior air,  tells $A$ that it
is he who has given $A$ and his notebook and even his memory of noting
something one week ago in the notebook their objective existence.  A
however does not take kindly to these assertions of superiority of $B$
and tells him not to feel so superior since, who knows, the objective
existence of $B$ itself may depend on some future observer as well.
\end{enumerate}

This example brings out the role of consciousness also.  The
suggestive nature of these examples partly derives from an appeal to
concepts like living-dead, consciousness etc.  All the same one
wonders as to why all these examples appear paradoxical if quantum
mechanics is the true description of nature?
\bigskip

\noindent {\large\bf 3. Rules of the Game}
\medskip

Before we proceed further to discuss various possible models of
quantum reality, it may be worthwhile to spell out the common ground
on which most physicists would agree.  There are the rules of the game
which one uses to make predictions for quantum phenomenon using various
relevant observational setups.

The state of a system $S$ is represented by a state vector
$|\psi\rangle$ belonging a Hilbert space $H$.  Every physical
observable $A$ is represented by a corresponding Hermitian operator
$\hat A$ in the Hilbert space $H$.  The expectation value of the physical
observable $A$, when the system is in the state $|\psi\rangle$, is
given by $\langle \psi|\hat A|\psi\rangle$.

The dynamics is given by the Schr\"odinger equation
\[
i\hbar {\partial\psi (t) \over \partial t} = H \psi (t)
\]
where $H$ is the Hamiltonian operator of the system $S$.  Let the
eigenvalues of the Hermitian operator $\hat A$ be denoted by $a_j$,
with the corresponding eigenvector $|a_j\rangle$, i.e.
\[
\hat A |a_j\rangle = a_j |a_j\rangle .
\]
Any measurement of the observable $A$ always yields one of eigenvalues
$a_j$ as the result of measurement.  If the system is in a state
$({\displaystyle \sum_n} |c_n|^2 = 1)$,
\[
|\psi\rangle = \sum_n c_n | a_n\rangle
\]
then a measurement of the observable $A$ will yield the value $a_n$
with the probability equal to $|c_n|^2$.
\bigskip

\noindent {\large\bf 4. Quantum Ontology}
\medskip

\nobreak
We discuss some of the various models of Quantum reality which have
been proposed.  We can classify them under two classes.  In the first
class we include those interpretations of quantum mechanical formalism
has to be supplemented by extra ingredients.  We include here Copenhagen
group of interpretations (Bohr, Heisenberg, von-Neumann and others)
and de-Broglie-Bohm's causal interpretation.  The second class of
interpretations of quantum mechanics are those where Quantum mechanics
does not need any extra supplementary ingredients.  This class
includes Everetts relative-state interpretation and related many
world or many-mind approaches.  It also includes the Quantum History
approach. 

\newpage

\noindent {\large 4.1. Quantum Ontology -- Class I}
\medskip

\nobreak
\noindent {\it 4.1.1. Copenhagen group of Interpretations}
\medskip

\nobreak
\noindent {\it 4.1.1.1. Niels Bohr}
\medskip

\nobreak
Niels Bohr insisted on the logical priority, as opposed to indubitable
chronological priority of classical physics.  According to him, the
language of classical Physics is the only one available to us 
to communicate observed results of any phenomenon.  The measuring devices
are to be described by classical physics.  With him this is a matter of
principle.  We are thus supplementing the formalism of quantum
mechanics by the addition of classical physics to describe measuring
apparatus.  For Bohr the combined system $(S)$ and measuring
apparatus $A$ is an unanalysable whole phenomenon $(S+A)$.  Thus
$(S+A_1)$, e.g. electron $S$ and it's position recorder $A_1$, and
$(S+A_2)$, e.g. electron $S$ and a momentum measuring device $A_2$,
are two different phenomenon.  They can not be combined into a single
description which would correspond to an electron $S$ with a known
position and momentum.  This is the essence of his complementarity
principle.  The different descriptions $(S + A_1)$ and $(S + A_2)$ are
complementary.  This is unlike the tale of ``the elephant and six blind
men'' where different descriptions can be combined into the
description of an ``elephant'' with preexisting attributes.  The
quantum world is probabilistic and a measurement does not reveal any
pre-existing properties of the system.

As Bohr said ``strictly speaking the mathematical formalism of quantum
mechanics and electrodynamics merely offer rules of calculation for
the deduction of \underbar{expectations} about observations obtained
under well-defined experimental conditions specified by classical
physical concept''.  Bohr thus had an ascetic attitude of renunciation
of any further picture of physical phenomenon.  One may say that for
him there was no ontology of quantum world.   

Bohr's solution of EPR problem was the system consisting of two
electrons can not be regarded as two ``separate'' electrons.

Mermin, in his Ithaca interpretation, says ``Correlations have
physical reality, that which they correlate does not''.  He contrasts
it with the ``another'' concept in electrodynamics.  ``Fields in empty
space have physical reality; the medium that supports them does not''.
\bigskip

\noindent {\it 4.1.1.2. Werner Heisenberg}
\medskip

\nobreak
Bohr did not assign any particular significance to the state vector
(or wavefunction) $\psi (t)$ of the quantum system, in his ``No
Ontology'' outlook.  In Heisenberg ontology, the $\psi(t)$, state
vector of 
the system, together with all the observables $O(t)$, represents, like
Aristotle's \underbar{potentia}, ``objective tendencies'' for any
``actual event'' to occur.  The state vector $\psi (t)$ represents the
wave like aspects of nature.

The particle like aspects are represented, in Heisenberg ontology, by
``actual events''.  It is here that the concept of ``the collapse of
the wavefunction'' is used.  The actualisation of the potentialities
happens when the experimental result is recorded e.g. the blackening
of the photographic plate, a click in a Geiger counter.  It could even
be a million year old fission track seen in a crystal.  As Heisenberg
said ``The observation itself changes the probability function
discontinuously; it selects of all possible events the actual one that
has taken place ..... the word ``happen'' to apply only to
observation, not to the state of affairs between observations, .....;
it is not connected with the act of registration of the result in the
mind of the observer''.

Karl Popper's use of ``propensitics'' is quite similar to Heisenberg's
use of ``potentia''.
\bigskip

\noindent {\it 4.1.1.3. Von-Neumann}
\medskip

\nobreak
In contrast to Bohr, the measuring apparatus $A$ as well as system $S$
are both to be described by quantum mechanics.

Let the eigenfunctions of the system $S$, corresponding to some
observable $\Omega$, be denoted by $\psi_n$, i.e. $\Omega \psi_n =
\omega_n \psi_n$, corresponding to an eigenvalue $\omega_n$.
Similarly let the pointer states of the apparatus be given by $Mf(a) =
af(a)$, where $M$ is the pointer position operator with eigenvalues
$a$ and wavefunction $f(a)$.  Von-Neumann postulates the measurement
interaction, which causes the initial system-apparatus state
\[
\psi_n f(a) \ {\rm to \ evolve \ into} \ \psi_n f(a_n).
\]
In view of superposition principle of quantum mechanics, if the system
is initially in a superposition ${\displaystyle \sum_n} c_n \psi_n$,
then the measurement interaction will cause the state
\[
\sum_n c_n \psi_n f(a) \ {\rm to \ evolve \ into} \ \sum_n c_n \psi_n
f(a_n). 
\]
This results in the probability of observing the pointer reading to be
$a_n$, corresponding to system being in the state $\psi_n$, to be
$|c_n|^2$. 

The state $\sum c_n \psi_n f(a_n)$ is a linear superposition of states
with different pointer readings.  It is a grotesque state.  We have
still not obtained a difinite value of the pointer reading.
Von-Neuman, now postulates, that when the measurement
of completed, and not before that, the wave function collapses to one
of the terms in the linear superposition e.g. to $\psi_N f(a_N)$.  It
is this collapse postulate which adds an extra ingredient to Quantum
mechanics and making quantum mechanical description nonclosed.

In an elaboration of von-Neuman view by London and Bauer, and also
subscribed to by Wigner and Stapp, this final collapse of wave
function takes place when the result of measurement is recorded in a
human mind.  As Stapp says ``There have been many efforts over the
intervening 70 years to rid physics of this contamination of matter by
mind ... . (I) claim that these decontamination efforts have
failed.''  Further he notes ``Nature is built not out of matter, but
out of knowledge''.
\bigskip

\nobreak
\noindent {\it 4.1.2 de Broglie-Bohm Causal Interpretation}
\medskip

\nobreak
Louis deBroglie proposed a realistic causal interpretation of quantum
mechanics in 1927 in his pilot wave theory. Due to dominance of
Copenhagen interpretation, criticism by Pauli and others, and 
lack of endorsement even by Einstein, who was against the Copenhagen
interpretation, this interpretation did not find general
acceptance. Even deBroglie gave it up. Bohm, in 1952, came up
independently with a similar proposal. He was also able to take care
of various objections against it. Further he supplemented it with a
theory of measurement. Recent interest was sparked by Bell's advocacy 
of it. 

The ontology of deBroglie-Bohm version is realistic. In this theory a
quantum object, say an electron, has both a particle aspect, it has a
trajectory $q(t)$ associated with it, as well as wave aspects, as its
wavefunction $\psi (q,t)$ is involved in determining it's velocity
$\dot q(t)$ at any time $t$ through the relation
\[
m \dot q (t) = \nabla S
\]
where $S$ is the phase of the wavefunction $\psi$. The time evolution
of wavefunction $\psi$ is given by the Schr\"odinger equation. The
quantum formalism here is supplemented by the addition of particle
trajections. 

The deBroglie-Bohm dynamics is deterministic. Give position $q(t)$ and
wavefunction $\psi (t)$ at $t=0$ we can predict it for all later
times. How do quantum probabilities arise? They do so through our
inability to precisely control particle positions. All we are capable
of is to prepare an statistical ensemble at $t=0$ in which particle
position $q$ is distributed according to the probability distribution
\[
P(q, t=0) = |\psi (q, t=0)|^2 .
\]
Once one has done that then deBroglie-Bohm dynamics makes sure that
the probability distribution $P(q,t)$ of particle positions at later
times is distributed according to 
\[
P(q,t) = |\psi (q,t)|^2 .
\]

The resolution of usual wave-particle duality conundrum here is not
that it is {\it particle or wave} according to the experimental set up
but thatit is both a particle and a wave. In the double slit
experiment the particle goes through one of the two slits but the wave
goes through both. 

It should not be concluded that we have reverted to classical
deterministic physics. The electron trajections are quite different
here as compared to classical physics e.g. in the double slit set the
trajectonis never cross mid plane between the slits. There is also
Einstein nonlocality as the trajectories are affected by the
wavefunction which depends even on far away potentials and/or boundary
condition on it. 
\bigskip

\noindent {\large 4.2. Quantum Ontology -- Class II}
\medskip

\nobreak
\noindent {\it 4.2.1 Everett}
\medskip

\nobreak    
Everett, in 1957, was first to propose an interpretation, in which the
Schr\"odinger equation provides a complete description of the
nature. Unlike Copenhagen interpretation the measuring apparatus has
to be described by quantum mechanics itself. We demand Schr\"odinger
equation describe the entire evolution. Unlike Heisenberg and
von-Neumann there is no ``collapse of the wavefunction'' in this
approach. It is a unitary description without any split between system
and apparatus. B. deWitt commented and popularised this interpretation
and renamed it as ``The many-world interpretation'' of quantum
mechanics. 

In general as a result of time evolution due to Schr\"odinger
equation, an observer would be expected to be in a superposition of
wavefunctions describing different eigenstates. If e.g. we are in a
normalised state 
\[
= c_1 \psi ({\rm walking~east}) + c_2 \psi ({\rm walking~west}) ,
\]
as a result of the interaction we should be simultaneously
walking to the east and to the west. Everett asserts that this is what
actually happens. As a result of measurement interaction we split
into two editions of ourselves and one of these walks to the east and
the other walks to the west. The universe has split. The universe in
which we are walking east has the probability equal to $|c_1|^2$ and
the universe in which we walk to the west has the probability equal to
$|c_2|^2$. 

How is it that we are not aware of this split? We seem to inhabit only
one universe. This situation has been contrasted with our not being
aware of any motion of the earth while Copernicus did assert that the
earth is moving. Copernicus had to explain the feeling of no motion by
pointing out that our relative motion with respect to the earth is not
there. The theory has to be only self consistant. Here also it is
asserted that we will not feel the split and that the probability
prescriptions used normally are natural to it. 

The theory maintains objective reality but at the cost of uncontrolled
multiplication of the universes. The universe splits every time an
observation is made. The wavefunction never collapses as an
observation is made. The different components of it inhabit different
universes. 

This interpretation has a certain fascination of it's own. Jorge Louis
Borges has even used such a possibly in one of his wonderful short stories. Yet
most people find it extravagant and wasteful of universes. In
contrast however it has a wonderful economy of principles. Only
Schr\"odinger equation and nothing else is need. 

Recent variants of this interpretation are ``many-mind'' instead of
``many-world'' interpretations of Albert and Loewer, Lockwood, Zeh and
others. 

This interpretation has been found to have particular appeal by those
working in the field of Quantum Computation. As pointed out by Deutch,
the massive parallel-computing associated with quantum computation is
easily understood for a superposition 
\[
\psi = \sum c_j \psi_i ,
\]
if we regard that a computation on each $\psi_i$ is carried out in a
different universe. 
\bigskip

\nobreak
\noindent {\it 4.2.2 Quantum History Approach}
\medskip

\nobreak    
This approach could be regarded as the minimalist completion of
Copenhagen approach to provide a closed system. It is associated
mainly with the work of R. Griffith, R. Omnes, M. Gell-Mann and
J.B. Hartle. 

The basic ingredients of the approach are:
\begin{enumerate}
\item[{(1)}] A possible set of fine grained histories $h$\\
There could be, e.g., a set of particle trajectory coordinates
$x(t=0), x(t_1), x(t_2), \cdots , x(t_N)$ at times $t=0, t_1, t_2,
\cdots , t_N$ where $0 < t,_1 < t_2 < \cdots < t_N$. In general a set
of projection operators at various times are used specify a find
grained history. 
\item[{(2)}] A notion of coarse graining \\
In the example about this could for example be taken to be $x (t_1)
\epsilon \Delta_j, x(t_2) t \Delta_j$, etc. where $\Delta_j$ are
various intervals to which $x(t_i)$ could belong. 
\item[{(3)}] A decoherence functional $D(h,h')$ \\
It is to be defined for each pair of histories $h$ and $h'$. The
functional $D(h, h')$ should satisfy 
\begin{eqnarray}
D(h, h') &=& D^\ast (h', h) \nonumber \\
D(h,h) &\geq& 0 \nonumber \\
\sum_{h,h'} D(h, h') &=& 1 . \nonumber
\end{eqnarray} 
\item[{(4)}] Superposition \\
For coarse graining we must have 
\[
D(\bar h, \bar h') = \sum_{h \subset \bar h} \sum_{h' \subset \bar h'}
D (h,h') .
\]
\item[{(5)}] 
The twohistories $h$ and $h'$ are said to decohere if
\[
\Re D(h,h') = 0 .
\]
For a set of decoherent alternative histories, we defind the
probability $p(h)$ for the history $h$, to be 
\[
p(h) = D (h,h) .
\]
This approach does not require observers for its formulation. As such
it is applicable to quantum cosmology. 
\end{enumerate}    

The interpretation provides for quantum probabilities only for
histories in a decoherent sets. ``Each set is part of
a complete quantum description of the system, but there is no
fine grained  decoherent set histories of which all the decoherent
sets are coarse grainings''. 

The values we observe for a physical observable are unlike Bohr and
others, are preexisting ones. That is a bit of realism here. It is
also asserted that among all decoherent sets there is also those with
a {\it quasi classical} realm. This is necessary for a quasi classical
description to be applicable to macroscopic everyday objects. 

Coming back to ontology, we have, as put by Hartle, ``Many {\it
complementary} descriptions of the universe that are mutually
incompatible''.
\bigskip

\noindent {\large\bf 5. Concluding Remarks}
\medskip

We thus see that Classical Physics had a straightforward and definite
ontology.  It was realistic.  This is quite unlike the discussions of
reality by classical and later philosophers, who could subscribe to an
spectrum of views ranging from `realism', `empiricism', and others
right upto ``idealism''.  They could even be totally skeptic as well.

The quantum physics has a well defined formalism to predict the
results of possible measurements.  However this formalism is quite fluid,
unlike classical physics, as far as it's ontology is concerned.  One
could refuse to talk of reality with Bohr, consider a ``poentia vs
actual'' state of affairs with Heisenberg or of ``mind and knowledge''
with von-Neumann and Wigner.  The deBroglie-Bohm picture of the
quantum world is in accord with realism but with many nonclassical
features.  You can subscribe to many-world/many-mind description or a
set of multiple alternative history description which are separately
consistent.  To use a phrase of Heinz Pagels, we have a veritable
``Reality Market place''.

In contrast to the ontological fluidity of the formalism of quantum
mechanism, the formalism itself appears to be exceptionally rigid.
Any attempt to tinker with it in any way leads to nonconservation of
probability, faster than light signalling or some other pathology.
Examples of modifications which have been tried are (i) addition of
small nonlinearities in Schr\"odinger equation (ii) the position
probability density $P$ is changed from $P = |\psi|^2$ to $P=
|\psi|^{2+\epsilon}$ with an small $\epsilon$.  It would appear that
Quantum Mechanics is an isolated point in the space of physical
theories.

At present among those interpretations which regard it as a unitary
description of the world the most developed one appears to be the
``quantum History approach''.  As a result of this development it is
able to encompass many more situations, including quantum cosmology
than other formalisms.
\bigskip

\begin{center}
\underbar{Acknowledgement}
\end{center}
\medskip

The author wishes to thank Prof. B.V. Sreekantan for inviting me to
deliver this talk.  Some portions of this talk are adapted from a 
earlier talks given at Ahemedabad and Indore.

\newpage

\begin{center}
\underbar{Bibiliographical Notes}
\end{center}
\medskip

\begin{enumerate}
\item[{}] The discussion in the present paper partially overlaps the
author's earlier reviews, 

Singh, V., Foundations of Quantum Mechanics, in \underbar{Gravitation,
Quanta and the universe}, ed. by A.R. Prasanna et al, Wiley Eastern,
New Delhi (1980).

Singh, V., Quantum Physics, Some Fundamental Aspects, in
\underbar{Schr\"odinger Centenary} \underbar{Surveys in Physics}, edited by
V. Singh and S. Lal, World Scientific, Singapore (1988).

\item[{1.}] Newtonian ontology is discussed in \\ Burtt, E.A.,
\underbar{The Metaphysical Foundations of Modern} \underbar{Science},
Humanities 
Press, Atlantic Highlands (1952), \\ For development of classical
physics, we refer to \\ Whittaker, E.T., \underbar{A history of
the theories of Acther and Electricity, Vol. 1 Classical} \hfill\break
\underbar{Theories}:
Harper Torchbacks (1960).  \\ For an overview of classical physics, see

Bergman, P.G., \underbar{Basic Theories of Physics;} \underbar{Mechanics and
Electrodynamics}, Prentice Hall, New York (c. 1949).

\item[{2.}] For a historical account of Quantum mechanics see,

Jammer, M., \underbar{The Conceptual Development of Quantum
mechanics}, Mc-Graw Hill, New York (1966) and \underbar{The Philosophy
of Quantum mechanics}, John Wiley, New York (1974).

Mehra, J. and Rechenberg, H., \underbar{The Historical Development of
Quantum mechanics}, Springer (1982 -- ~~~~).

Whittaker, E.T., \underbar{A theory of Acther and Electricity, Vol. 2,
Modern Theories} (1900-1926), Harper Torchbacks (1960).

Beller, M., Quantum Dialogue, Chicago (199).

\item[{3.}] (i) A number of important basic sources on fundamental
aspects of quantum mechanics are collected in 

Wheeler, J.A. and Zurek, W.H., \underbar{Quantum Theory and
Measurement}, Prinception Univ. Press (1983).

(ii) The writings of John Bell on quantum mechanics, and these are a
must for anyone interested in this area, are available in the
collection, 

Bell, J.S., \underbar{Speakable and unspeakable in quantum mechanics},
Cambridge (1987).

\item[{4.}] Some popular level books on quantum mechanics are 

Pagels, H.R., \underbar{The cosmic code : Quantum Physics as the
Language of Nature}, Pengis Books (1984).

Polkinghorn, J.C., \underbar{The Quantum World}, Longmans, London
(1984).

Rae, A., \underbar{Quantum Physics: Illusion or Reality}, Cambridge
Univ. Press (1986).

A more advanced text is

deEspagnat, B., \underbar{Conceptual Foundations of Quantum
Mechanics}, Benjamin (1976).

\item[{5.}] (i) The best descriptions of double slit experiments is by
Feynman, R.P., in  

Feynman, R.P., \underbar{The Feynman Lectures on Physics}, vol. 3,
Addison-Wesley, (1966).

(ii) For Wheeler's Delayed Choice Double slit experiment, see

Wheeler, J.A., The `past' and `delayed choice' double slit
experiment, in \underbar{Mathematical} \underbar{Foundations of Quantum
Mechanics}, edited by Harlow, A.R., Academic, New York (1960).

(iii) The role of uncertainty principle in upholding consistency of
quantum behaviour of electrons is vividly brought out in the classic 

Bohr, N., Discussions with Einstein on Epistomological Problems in
Atomic Physics in \underbar{Albert Einstein, Philosopher-Scientist},
ed. by Schipp, P.A., Evanston (1949).

(iv) For developments in neutron interferometry we refer to

Greenberger, D.M., The neutron interferometer as a device for
illustrating the strange behaviour of quantum systems,
Rev. Mod. Phys. \underbar{55}, 875 (1983).

\item[{6.}] For a survey of hidden variable theories we refer to

Belinfante, F.J., \underbar{A Survey of Hidden Variable Theories},
Pergamon, Oxford (1973).

One should of course read also the masterly treatment in 

Bell, J.S., \underbar{On the Problem of Hidden Variables in Quantum
Mechanics}, Rev. Mod. Phys. \underbar{38}, 447 (1966).

\item[{7.}] (i) The original E-P-R paper appeared in 

Einstein, A., Podolsky, B. and Rosen, N., Phys. Rev. \underbar{47},
777 (1935). 

The restatement quoted is from the volume edited by 

Schilpp, P.A., \underbar{Albert Einstein : Philosopher-Scientist},
Library of Living Philosophers, Evanston III (1940).

(ii) Bohm's version appears in 

Bohm, D., \underbar{Quantum Theory}, Prentice Hall (1951).

(iii) For E.P.R. Paradox and Bell's theorem, see

Redhead, M., \underbar{Incompleteness, Nonlocality and Realism},
Clarendon Press, Oxford (1987).

Selleri, F., \underbar{Quantum Mechanics versus Local Realism: The
Einstein-Podolsky-Rosen} \hfill\break \underbar{Paradox}, Plenum (1988).

Cushing, J.T. and Mc Mullin, E., \underbar{Philosophical consequences
of Quantum Theory:} \hfill\break \underbar{Reflections on Bell's
Theorem}, Notre Damn Press (1989).

Selleri, P., \underbar{Quantum Paradoxes and Physical Reality},
Kluwer, Dordrecht (1990).

Home, D. \underbar{Conceptional Foundations of Quantum Physics},
Plenum, New York (1997).

(iv) For Hardy's version of EPR paradox, see 

Hardy, L.,
Phys. Rev. Lett. \underbar{68}, 2981-2984 (1992), 

Boschi, D. et. al.,
Phys. Rev. Lett. \underbar{79}, 2755-2758 (1997).

\item[{8.}] (i) Schr\"odinger's cat was discussed first in

Schr\"odinger, E., Naturwissenschaften \underbar{23}, 807-12, 823-828,
844-849 (1935).

An english translation is available in Wheeler and Zurek.

(ii) Wigner's friend was discussed first in 

Wigner, E., ``Remarks on the
Mind-Body Question'' in Good (ed). \underbar{The Scientist
Speculates}, Heineman, London (1961).

\item[{9.}] (i) For Bohr's views we refer to

Bohr, N., \underbar{Atomic Theory and the Description of Nature},
Cambridge (1934); 

Bohr, N., \underbar{Atomic Physics and Human Knowledge}, Wiley, New York
(1958); 

Bohr, N., \underbar{Essays 1958/1962 on Atomic Physics and Human
Knowledge}, Wiley, New York (1963); 

(ii) Heisenberg, W., Daedalus \underbar{87}, 95-108 (1958),

Heisenberg, W., \underbar{Physics and Philosophy}, Harper, New York
(1958).  

Popper, K., Brit. J. Phil. Sc. \underbar{10}, 25-42 (1959).
 
(iii) J. von-Neumann's theory of measurement is contained in his book

von-Neumann, J., \underbar{Mathematische Grundlagen der Quanten
Mechanik}, Springer Verlag, Berlin (1932), \underbar{Mathematical
Foundations of Quantum mechanics}, (trans. by R.T. Beyer), Princeton
(1955). 

(iv) London and Bauer's elaboration is contained in their book 

London, F. and Bauer, E., \underbar{La Theorie de L' observation en
Mecanique Quantique}, Hermann and Cie, Paris (1939).

An English translation is available in the collection by Wheeler and
Zurek referred earlier.

(v) Stapp, H., \underbar{Mind, Matter and Quantum Mechanics},
Springer-Verlag, (1993). 

\item[{10.}] The original papers on deBroglie-Bohm theory are

deBroglie, L., J. Physique, 6th series \underbar{8}, 225 (1927),

D. Bohm, Phys. Rev. \underbar{85}, 166-193 (1952),

D. Bohm, Phys. Rev. \underbar{89}, 458-466 (1953).

For a detailed treatment, see

Bohm, D. and Hiley, B.J., \underbar{The Undivided Universe},
Routledge, London (1993), 

Holland, P.R., \underbar{Quantum Theory of Motion}, Cambridge (1993)

Cushing, J.T., \underbar{Quantum Mechanics -- Historical Contingency
and Copenhagen} \hfill\break \underbar{Hegemony, Chicago}, (1994).

For some recent developments see, 

Roy, S.M. and Singh, V., Phys. Letters \underbar{A255}, 207-208
(1999),

D\"urr, D., Goldstein, S., Tumulka, R. and Zanghi, N.,
Phys. Rev. Lett. \underbar{93}, 090402-1 (2004).

\item[{11.}] For relative-state interpretation related interpretations  
see

Everett, H., Rev. Mod. Phys. \underbar{29}, 454 (1964), 

deWitt, B.S. and Graham, N., \underbar{Many World interpretation of
Quantum Mechanics}, Princeton (1973),

Lockwood, M., ``Many Minds'' Interpretation of Quantum Mechanics,
Brit. J. Phil. Sc. \underbar{47}, \#2 (1996).

Deutsch, M., \underbar{The Fabric of Reality}, Penguin (1997).

\item[{12.}] For quantum History approach see,

Griffith, R.B., Stat. Phys. \underbar{36}, 219-272 (1984),

Griffith, R.B., \underbar{Consistent Quantum Theory}, Cambridge (2002),

Gell-Mann, M. and Hartle, J., Phys. Rev. \underbar{D47}, 3345-3382
(1993).

Hartle, J., \underbar{The Quantum mechanics of cosmology}, Jerusalem
Lectures (1989-90).

M. Gell-Mann, \underbar{The Quark and the Jaguar}, Abaus (1994).

R. Omnes, \underbar{The Interpretation of Quantum Mechanics},
Princeton (1994).

Zurek, W.H., Rev. Mod. Phys. \underbar{75}, 715-775 (2003).

\item[{13.}] S. Weinberg, Ann. Phys. (NY) \underbar{194}, 336 (1989).

S. Weinberg, Phys. Rev. Lett. \underbar{62}, 485 (1989).

Aaronson, S., arXiv: quant-ph/0401062 (2004).

For a listing of works on nonlinear modifications of Schr\"odinger
equation see

Svetlichny, G., arXive: quant-ph/0410036, (2004).

\item[{14.}] References for the various quotes are provided here
except for those from the references already cited.  The Boltzmann
quote from Goethe in Sec. 1.2 is from 

Herman A.,  \underbar{The New Physics}, Internationes, Bonn (1979) p.7-8.  

Minkowski's statement is
from an address he gave to eighteeth assembly of German natural
Scientists and physicians at Cologne on 21 Sept. 1908.  

The quote from
d'Espagnat about atonism is from his lecture, ``Quantum Physics and
Ontological Problem'', available on the web.  

The quotations from
Stapp are from his paper `Quantum Ontology and Mind-Matter Synthesis'
(LBNL-40722, 1988).  

In Sec. 4.2.2 I am unable to trace the source of
the quotation ``Each set ..... coarse grainings''.  

The quote from
Hartle is from his paper ``What Connects Different Interpretations of
Quantum Mechanics'' (arXiv: quant-ph/0305089, 2003).
\end{enumerate} 
\end{document}